# Some Observations on Optimal Frequency Selection in DVFS–based Energy Consumption Minimization

Nikzad Babaii Rizvandi[1,2], Javid Taheri[1], and Albert Y. Zomaya[1]

[1] Centre for Distributed and High Performance Computing
School of Information Technologies, University of Sydney
NSW 2006, Australia

[2] National ICT Australia (NICTA), Australian Technology Park
Sydney, NSW 1430, Australia

nikzad@it.usyd.edu.au
{javid.taheri, albert.zomaya}@sydney.edu.au

**Abstract**

In recent years, the issue of energy consumption in parallel and distributed computing systems has attracted a great deal of attention. In response to this, many energy−aware scheduling algorithms have been developed primarily using the dynamic voltage−frequency scaling (DVFS) capability which has been incorporated into recent commodity processors. Majority of these algorithms involve two passes: schedule generation and slack reclamation. The former pass involves the redistribution of tasks among DVFS−enabled processors based on a given cost function that includes makespan and energy consumption; and, while the latter pass is typically achieved by executing individual tasks with slacks at a lower processor frequency. In this paper, a new slack reclamation algorithm is proposed by approaching the energy reduction problem from a different angle. Firstly, the problem of task slack reclamation by using combinations of processors' frequencies is formulated. Secondly, several proofs are provided to show that (1) if the working frequency set of processor is assumed to be continues, the optimal energy will be always achieved by using only one frequency, (2) for real processors with a discrete set of working frequencies, the optimal energy is always achieved by using at most two frequencies, and (3) these two frequencies are adjacent/neighbouring when processor energy consumption is a convex function of frequency. Thirdly, a novel algorithm to find the best combination of frequencies to result the optimal energy is presented. The presented algorithm has been evaluated based on results obtained from experiments with three different sets of task graphs: 3000 randomly generated task graphs, and 600 task graphs for two popular applications (Gauss-Jordan and LU decomposition). The results show the superiority of the proposed algorithm in comparison with other techniques.

## 1. INTRODUCTION

Research on low power systems has been attracting a great deal of attention in recent years across a number of areas and technologies. A few examples of these systems are:

- Wireless sensors: several sensors extract data from the environment, transmit these data to a processing unit and receive processed data accompanied by appropriate commands from the processing unit [1]. The sensors and their receiver/transmitter are generally powered by battery and/or solar cells.
- Satellite circuits: Satellites typically contain massive number of complex circuits which must work at low power. These circuits are supplied by solar cells, the only available power supply in satellites.
- Robots and surveillance devices: these devices are heavily used in army, mine extraction and in unsafe environments for humans.
- Cell phones and laptops: these devices are powered by batteries which are expected to work for prolonged periods of time.

In recent years, the high price of energy and a variety of environmental issues have forced the high performance computing sector to reconsider some of its old practices with an aim to create more sustainable HPC systems. The Earth Simulator with a power consumption of 12 MW/h and Petaflop with a power consumption of 100 MW/h are two typical examples of such energy hungry HPC systems [2-3]. The magnitude of this consumption will be even greater if the energy consumption of the associated cooling systems is also considered. For example, the survey in [4] indicates that if number of transistors in processor –1 billion for the recent Intel Itanium 2–continues with the current increasing rate, produced heat (per $cm^2$) by future processors will exceed the sun's surface temperature.

Therefore, new processor architectures require mechanisms that reduce energy consumption so that the amount of emitted heat can also be reduced [5-6]. Furthermore, not only does the rising temperature of a circuit derail its performance, but it can also lead to significant shortening of the lifetime of its components. For example, a formula based on Arrhenius Law indicates that lifetime expectancy of many HPC systems components is halved for every $10^o C$ temperature increase [3].

To reduce energy consumption in HPC systems, or clusters on a smaller scale, resource management in both hardware and software must be addressed. One issue in hardware resource management—which has direct dependency on the number of transistors—is to efficiently reduce energy consumption by processors. Dynamic voltage–frequency scaling (DVFS), already incorporated into many recent processors, is perhaps the most appealing method for reducing energy consumption. DVFS reduces energy consumption of processors based on the fact that such energy consumption in CMOS circuits has a direct relationship with (1) working frequency and (2) the square of the supplied voltage. Thus, DVFS saves energy by switching between processor's voltages/frequencies to execute tasks during *slack* times. Although DVFS was originally designed for task scheduling on single processors [4, 7-10], however, it has recently been extended and used in parallel and distributed computing systems as well [3, 11].

To deploy DVFS, it must be properly integrated with a task scheduler by using one of the following two approaches: (1) during the scheduling process or (2) slack reclamation after scheduling. In the first approach, tasks graph are scheduled on DVFS−enabled processors by minimizing both energy and makespan at the same time [12-13]. In the second approach, an independent scheduler is first used to distribute tasks among processors without considering energy consumption. This procedure is then followed by an independent DVFS technique to minimize energy consumption of tasks by filling the generated tasks' slack times.

The existing methods based on DVFS techniques, however, have two major limitations: (1) most of them still focus on the scheduler and rarely explore other opportunities for slack reclamation, and (2) they only use one frequency (among a discrete set of frequencies) to perform each task—the use of one frequency usually results in underutilized slack times leading to energy wastage by processors and other devices.

In this paper, we propose a new slack reclamation technique to reduce energy consumption of processors through efficient use of the generated tasks' slack times by an independent scheduler. In our approach, hereafter called Multiple Voltage-Frequency Selection DVFS (MVFS−DVFS), the key idea is to execute tasks using a linear combination of available frequencies so that all slack times are fully utilized. The MVFS−DVFS algorithm is presented in three steps. Firstly, energy consumption of each task is formulated as an optimization problem with constraints. Secondly, formal proofs are provided to show that the optimal set of at most two voltage-frequencies will always lead to minimum energy consumption. Also, if power consumption is modelled as a convex function of frequency, these two frequencies are adjacent. Thirdly, an algorithm is proposed to find these aforementioned frequencies for each task. Performance of MVFS−DVFS is compared against previous approaches with similar goals. The rest of the paper is organized as follows. Section 2 presents related works. Section 3 describes preliminaries including our assumed system and energy models. In Section 4, MVFS−DVFS algorithm is presented. Experimental results and conclusions are presented in Sections 5 and 6, respectively.

## 2. RELATED WORK

In recent years there has been a significant amount of work on task scheduling for real−time embedded systems using various forms of DVFS enabled techniques. The main idea in most of the existing algorithms is to efficiently use processors' slack times to satisfy time requirements of all tasks; e.g. deadlines, release times and execution times. Based on provided/estimated information for each task, energy−aware task scheduling algorithms in embedded systems can be categorized into two groups: static (offline) and dynamic (online). In static scheduling timing information of all tasks is made available during compile−time, scheduling is performed to meet all deadlines while maximizing processor utilization [7], [8], [12], [14], [15]. This type of scheduling is used in most large−scale computational problems, such as, bioinformatics [16], chemistry [17] and machine vision applications [18]. In dynamic scheduling, on the other hand, although tasks' deadlines might be available during compile−time, their release and execution times must be estimated during the run−time [3], [10], [13], [19]. This class of scheduling is usually used in dynamic large−scale approximation and optimization problems such as weather forecasting [20] and search algorithms [21] as well as most power−aware devices like laptops, wireless sensors and cell phones. While there are many algorithms in the literature for energy−efficient both static and dynamic scheduling on uni−processor and multiprocessor systems, most of them cannot be applied to reduce energy consumption in clusters. In fact, these algorithms are suitable for systems with small number of processors [22-23] as well as those with shared memories [4, 15]. In addition, these algorithms mostly assume that the tasks (periodic or aperiodic) are independent [22-25].

Kappiah et al. in [26] used a just-in-time DVFS technique to fill slack times in MPI programs. A system called Jitter was utilized to reduce working frequency of nodes with more slack times and/or less assigned computation. Jitter ascertains that tasks would arrive just in time without increasing overall execution time. Ge et al. in [3] applied the DVS technique to processors that do not work at their peak performance during the execution of parallel applications. In this approach, the best processor frequency for each task was selected before its execution based on through analysis of collected computation and communication power profiles. A method to reduce energy consumption was presented in [27] to adaptively activate and deactivate hardware resources (e.g., memory) for intensive HPC applications.

Lee and Zomaya in [13] presented a DVFS−based algorithm to simultaneously minimize both completion time and energy consumption of precedence−constrained (dependant) parallel jobs. Their final result was a trade−off between quality of scheduling and consumption of energy. Ding et al. in [28] formally modelled efficiency/iso−efficiency concepts for energy scalability. They also extended their results to produce an analytical model for studying tradeoffs between performance and energy saving in HPC systems. Molnos et al. in [29] classified the slack times in real−time applications into static, work and shared lack groups for multiple dependent tasks on multiple DVFS−enabled processors. Then a dynamic dependency aware task scheduling was proposed to adjust voltage/frequency of the deadlines for tasks assigned to processors. Hotta [30] presented a profiling−based power−performance optimization method in which execution of a program was divided into several regions. In this approach, profile information for each region (including power and execution profiles) was extracted and then utilized to find the best combination of processor voltages and frequencies. In Springer et al. work in [31] an upper limit for system energy usage was first chosen externally; then, a combination of performance modelling and performance prediction was used to modify execution times according to this upper limit. After creating models for both execution time and energy consumption, key parameters of models were estimated by executing a program for a small number of times followed by regression. Here, for better estimation of parameters, the following steps were iterated until a proper schedule is achieved: (1) using models to predict each possible scheduling of tasks, (2) executing the program a few times with the best predicted schedule and (3) updating estimated key parameters. The use of multiple voltages in Dynamic Voltage Scaling enabled processors was used in Ishihara work in [32]. Their work is a simplified version of our work in this paper which will be described briefly in Section 4.4.

Kimura et al in [11] proposed an energy reduction algorithm for power−scalable high performance cluster supported by DVFS technique. This algorithm selects a suitable set of voltages and frequencies to execute tasks as uniformly as possible using the lowest available frequency with slightly increasing the overall execution time. In our former approach [33], an algorithm was proposed to reclaim slack times of tasks by linear combination of the processor highest and lowest frequencies. To the best of our knowledge, Reference DVFS algorithm (RDVFS) [11], and Maximum–Minimum–Frequency DVFS (MMF–DVFS) [33] are the most efficient algorithms with similar objectives to our work in this paper; therefore, they will be used to measure efficiency of our new approach.

## 3. PRELIMINARIES

This section describes the target system and application models and introduces the relevant energy models.

### *3.1. System and Application Models*

In this work, a parallel computing system is comprised of $N$ homogeneous processors with individual memories. In such systems, switching time between frequencies can be safely ignored in processors because time to switch from one frequency to another ($30-150\mu\sec$ [4]) is significantly smaller than execution time of tasks (at least $1m\sec$).

A set of dependent tasks, $\{A^{(1)}, A^{(2)}, ..., A^{(M)}\}$, represented by a directed acyclic task graph (DAG) is also assumed to be executed in the modelled HPC system. Here, the $k^{th}$–task ($A^{(k)}$) have the following four parameters: (1) $T^{(k)}$ as the whole available time a processor can assign to the task—summation the task's execution and slack time (Figure 1a), (2) $t_i^{(k)}$ as the task execution time when frequency $f_i$ is used, (3) $f_{ideal}^{(k)}$ as the ideal continuous frequency based on [34] that results the optimum energy consumption (Figure 1c), (4) $K^{(k)}$ as the required number of clock ticks (i.e. clock cycles) the task needs for its execution, and (5) $t_{OS}^{(k)}$ is the time the processor spends for executing the task in original scheduling (Figure 1a).

### *3.2. Energy Model*

DVFS–enabled processors can execute a task by using a discrete set of voltages–frequency pairs, $(f_i, v_i)$, in which $\{v_1 < v_2 < ... < v_N\}$ and $\{f_1 < f_2 < ... < f_N\}$. In CMOS based processors, the power consumption of a processor consists of two parts: (1) dynamic part that is mainly related to CMOS circuit switching energy, and (2) static part that addresses the CMOS circuit leakage power. The whole power consumption ($P_d$) is estimated as [4]:

$$\begin{cases} P_d = \lambda f v^2 + \mu v \\ \\ f \propto \dfrac{(v - v_t)^2}{v_t} \end{cases} \quad (1)$$

Where $f$, $\lambda$ and $v$ represent processor's working frequency, the effective capacitance, and processor's working voltage, respectively. Note that, $v_t$ is a threshold voltage usually provided by the manufacturer. In this paper, we consider a general relation between voltage, frequency and power as:

$$IF \quad (f_i, v_i) < (f_j, v_j) \quad THEN \quad P_d(f_i, v_i) < P_d(f_j, v_j) \qquad (2)$$

The overall energy consumption of $k^{th}$–task $(A^{(k)})$ in a DAG is calculated as:

$$E^{(k)} = P_d t_i^{(k)} + P_I (T^{(k)} - t_i^{(k)}) \qquad (3)$$

where $P_I$ is the energy a processor consumes when it is in idle.

## 4. MULTIPLE VOLTAGE–FREQUENCY SELECTION FOR DYNAMIC VOLTAGE–FREQUENCY SCALING (MVFS–DVFS)

In this section, the general DVFS problem is formally defined and our algorithm, Multiple Voltage–Frequency Selection Dynamic Voltage–Frequency Scaling (MVFS–DVFS), is provided.

### 4.1. Problem Statement

Optimal energy consumption of $k^{th}$–task can be defined as finding the best combination of available voltage–frequencies, $\{(f_1, v_1) < ... < (f_N, v_N)\}$ to perform a predefined task with $K$ clock ticks within a predefined time $T$. For the $k^{th}$–task, this optimal answer is defined as follows:

$$\begin{cases} Minimize \quad E^{(k)} = \sum_{i=1}^{N} t_i^{(k)} P_d(f_i, v_i) + P_I \left( T^{(k)} - \sum_{i=1}^{N} t_i^{(k)} \right) \\ s.t \\ \quad 1. \quad \sum_{i=1}^{N} t_i^{(k)} f_i = K^{(k)} \\ \quad 2. \quad \sum_{i=1}^{N} t_i^{(k)} \leq T^{(k)} \\ \quad 3. \quad t_i^{(k)} \geq 0; \; for \; i = 1, 2, \cdots, N \end{cases} \qquad (4)$$

Because our algorithm reclaims the slack time of each task independent from other tasks in DAG, the above formulation for the $k^{th}$–task is further simplified by replacing $t_i^{(k)}$, $T^{(k)}$ and $K^{(k)}$ with $t_i$, $T$ and $K$, respectively. Here, $t$ and $T$ are time values in miliseconds and $K$ is an integer value.

### 4.2. Computing the Optimal Solution

To find the optimal solution for the problem defined by equation (4), a simplified version of this problem is solved first; then generalized to find the solution for equation (4). This simplified version uses only three frequencies $((f_1, v_1) < (f_2, v_2) < (f_3, v_3))$ to perform a task in exact time (T) –oppose to within– and is defined as follows:

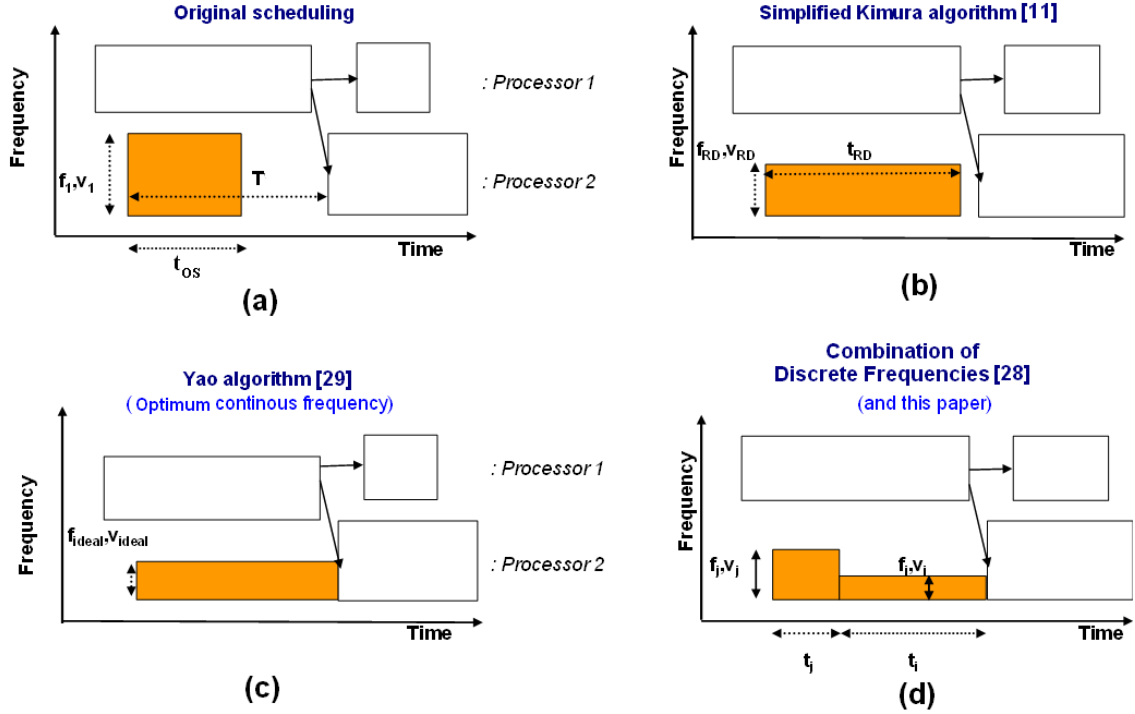

**Figure 1. Time representation of different DVFS−based algorithms for $k^{th}$−task.**

$$\begin{cases} Minimize \quad E = t_1 P_d(f_1, v_1) + t_2 P_d(f_2, v_2) + t_3 P_d(f_3, v_3) \\ s.t \\ \quad 1. \quad t_1 f_1 + t_2 f_2 + t_3 f_3 = K \\ \quad 2. \quad t_1 + t_2 + t_3 = T \\ \quad 3. \quad t_i \geq 0; \, for \, i = 1, 2, 3 \end{cases} \quad (5)$$

**Theorem 1:** The optimal solution for equation (5) is obtained by at most two voltage-frequencies.

**Proof:** To prove this theorem, the general energy formulation using three voltage-frequencies is first computed, and then minimized.

From constraints 1 and 2:

$$t_1 = T - (t_2 + t_3)$$
$$K = t_1 f_1 + t_2 f_2 + t_3 f_3 = (T - t_2 - t_3) f_1 + t_2 f_2 + t_3 f_3$$

$\Rightarrow$

$$t_2 = \frac{(K - Tf_1) - t_3(f_3 - f_1)}{f_2 - f_1} \quad (6)$$

$$t_1 = (T - t_2 - t_3) = T - \left[\frac{(K - Tf_1) - t_3(f_3 - f_1)}{f_2 - f_1} + t_3\right] = \frac{(Tf_2 - K) + t_3(f_3 - f_2)}{f_2 - f_1} \quad (7)$$

Based on constraint 3:

$$t_1 \geq 0 \Rightarrow (Tf_2 - K) + t_3(f_3 - f_2) \geq 0 \Rightarrow t_3 \geq \frac{K - Tf_2}{f_3 - f_2}$$

$$t_2 \geq 0 \Rightarrow (K - Tf_1) - t_3(f_3 - f_2) \geq 0 \Rightarrow t_3 \leq \frac{K - Tf_1}{f_3 - f_1}$$

$$t_3 \geq 0$$

which results in:

$$\begin{cases} 0 \leq t_3 \leq \dfrac{K - Tf_1}{f_3 - f_1} \\ \dfrac{K - Tf_2}{f_3 - f_2} \leq t_3 \leq \dfrac{K - Tf_1}{f_3 - f_1} \end{cases}$$

By replacing $t_1$ and $t_2$ in energy formulation based on $t_3$, the following equation for energy is obtained:

$$\begin{aligned} E &= t_3\left(\frac{(f_3 - f_2)P_d(f_1, v_1) - (f_3 - f_1)P_d(f_2, v_2) + (f_2 - f_1)P_d(f_3, v_3)}{f_2 - f_1}\right) \\ &+ \frac{(Tf_2 - K)P_d(f_1, v_1) + (K - Tf_1)P_d(f_2, v_2)}{f_2 - f_1} \\ &= \alpha t_3 + \beta \end{aligned} \quad (8)$$

This equation reveals that energy consumption of a task can be represented as a linear function of $t_3$. Depending on the sign of $\alpha$ two scenarios might arise: (1) $\alpha < 0$ (2) $\alpha > 0$.

Case 1: IF $\alpha < 0$; then, energy in equation (8) is a strictly decreasing function of $t_3$. Therefore, it is minimized when $t_3$ is set to its highest possible value. Thus:

$$\begin{cases} t_3 = \dfrac{K - Tf_1}{f_3 - f_1} \\ t_2 = 0 \\ t_1 = T - t_2 - t_3 = T - \dfrac{K - Tf_1}{f_3 - f_1} = \dfrac{Tf_3 - K}{f_3 - f_1} \\ E^*(f_1, f_3) = t_1 P_d(f_1, v_1) + t_3 P_d(f_3, v_3) = \dfrac{Tf_3 - K}{f_3 - f_1} P_d(f_1, v_1) + \dfrac{K - Tf_1}{f_3 - f_1} P_d(f_3, v_3) \end{cases} \quad (9)$$

Case 2: IF $\alpha > 0$; then, energy in Eqn. (8) is a strictly increasing function of $t_3$. Therefore, it is minimized when $t_3$ is set to its lowest possible value. Two minimal values might be set for $t_3$:

$$t_3 = 0 \Rightarrow \begin{cases} t_1 = T - t_2 - t_3 = T - \dfrac{K - Tf_1}{f_2 - f_1} = \dfrac{Tf_2 - K}{f_2 - f_1} \\ t_2 = \dfrac{(K - Tf_1) - t_3(f_3 - f_1)}{f_2 - f_1} = \dfrac{K - Tf_1}{f_2 - f_1} \\ E^*(f_1, f_2) = t_1 P_d(f_1, v_1) + t_2 P_d(f_2, v_2) = \dfrac{Tf_2 - K}{f_2 - f_1} P_d(f_1, v_1) + \dfrac{K - Tf_1}{f_2 - f_1} P_d(f_2, v_2) \end{cases} \quad (10)$$

or,

$$t_3 = \dfrac{K - Tf_2}{f_3 - f_2} \Rightarrow \begin{cases} t_1 = 0 \\ t_2 = \dfrac{(K - Tf_1) - \left(\dfrac{K - Tf_2}{f_3 - f_2}\right)(f_3 - f_1)}{f_2 - f_1} = \dfrac{Tf_3 - K}{f_3 - f_2} \\ E^*(f_2, f_3) = t_2 P_d(f_2, v_2) + t_3 P_d(f_3, v_3) = \dfrac{Tf_3 - K}{f_3 - f_2} P_d(f_2, v_2) + \dfrac{K - Tf_2}{f_3 - f_2} P_d(f_3, v_3) \end{cases}$$

(11)

Equations 9–11 show that regardless of whether energy is a strictly decreasing or increasing function of $t_3$, always two voltage–frequencies provide the optimal energy consumption.

**Corollary 1:** If two voltage–frequencies $(f_j, v_j) > (f_i, v_i)$ are capable of performing a task; then, their associated optimal energy consumption would be:

$$E^*(f_i, f_j) = \dfrac{Tf_j - K}{f_j - f_i} P_d(f_i, v_i) + \dfrac{K - Tf_i}{f_j - f_i} P_d(f_j, v_j) \quad (12)$$

**Proof:** direct observation from theorem 1.

**Corollary 2:** If two voltage–frequencies $(f_j, v_j) > (f_i, v_i)$ are capable of performing a task; then, their associated execution times for optimal energy consumption would be:

$$\begin{cases} t_i = \dfrac{Tf_j - K}{f_j - f_i} \\ \\ t_j = \dfrac{K - Tf_i}{f_j - f_i} \end{cases} \quad (13)$$

**Proof:** direct observation from theorem 1.

**Lemma 1 (Optimum continuous frequency):** If a processor is able to perform a task with a continuous range of voltage–frequencies, which is an unrealistic assumption, then the optimum energy to perform task $A$ is when task's slack time (T) is fully utilized.

**Proof:** If $(f_j, v_j) > (f_i, v_i)$ are two voltage–frequencies to obtain optimal energy for a task, then, equation (5) can be rewritten as:

$$\begin{cases} \text{Minimize} \quad E = t_i P_d(f_i, v_i) + t_j P_d(f_j, v_j) \\ s.t. \\ \quad 1. \quad t_i f_i + t_j f_j = K \\ \quad 2. \quad t_i + t_j = T \\ \quad 3. \quad t_i \geq 0, t_j \geq 0 \end{cases}$$

By replacing $t_j$ with $t_j = T - t_i$, the energy formula would be:

$$E = t_i \left( P_d(f_i, v_i) - P_d(f_j, v_j) \right) + P_d(f_j, v_j) \times T \tag{14}$$

Because $E$ in equation (14) is a strictly decreasing function of $t_i$, it is minimized when $t_i = 0$. This implies that if frequency can be chosen from a continuous spectrum, the energy is optimized using only one voltage–frequency. Further, this frequency would cover the whole slack time and could be calculated as follows:

$$t_i = 0 \Rightarrow \begin{cases} f_{ideal} = f_j = \dfrac{K}{T} \\ t_{ideal} = t_j = T \\ E_{Opt-Cont.} = T P_d(f_{ideal}, v_{ideal}) \end{cases} \tag{15}$$

**Lemma 2:** If a processor's set of available voltage-frequencies is discrete; then, two voltage-frequencies that would lead to the optimal energy consumption will be on both sides of $f_{ideal}$ in equation (15).

**Proof:** Constraint 3 in equation (5) implies that all time segments are greater or equal to zero. By applying this rule to the time values in Corollary 2 and with condition $(v_j, f_j) > (v_i, f_i)$, the following can be concluded:

$$\begin{cases} t_i \geq 0 \Rightarrow \dfrac{Tf_j - K}{f_j - f_i} \geq 0 \Rightarrow f_j \geq \dfrac{K}{T} \\ t_j \geq 0 \Rightarrow \dfrac{K - Tf_i}{f_j - f_i} \geq 0 \Rightarrow f_i \leq \dfrac{K}{T} \end{cases} \Rightarrow \quad f_i \leq \dfrac{K}{T} \leq f_j \quad (16)$$

where $\dfrac{K}{T} = f_{ideal}$ by definition.

**Theorem 2:** Optimal answer for equation (4) uses at most two voltage-frequencies.

**Proof (by contradiction):** To prove this theorem, we show that the optimal answer for equation (4) cannot use more than two voltage-frequencies to minimize total energy consumption. If we assume that the optimal answer for equation (4) utilizes more than two voltage-frequencies, then, its utilization profile can be depicted as that in Figure 2 for $n \geq 3$. In this case, this total task can be divided into two independent subtasks: (1) a subtask $(T_1, K_1)$ that uses three voltage-frequencies, e.g. $(v_n, f_n), (v_{n-1}, f_{n-1}), (v_{n-2}, f_{n-2})$ and (2) a subtask to cover the rest of calculations, i.e. $(v_{n-3}, f_{n-3}), \ldots, (v_1, f_1)$.

$$\begin{cases} T = T_1 + T_2 = (t_n + t_{n-1} + t_{n-2}) + (t_{n-3} + \ldots + t_1) \\ K = K_1 + K_2 = (t_n f_n + t_{n-1} f_{n-1} + t_{n-2} f_{n-2}) + (t_{n-3} f_{n-3} + \ldots + t_1 f_1) \\ E = E_1 + E_2 = E_1((v_n, f_n), (v_{n-1}, f_{n-1}), (v_{n-2}, f_{n-2})) + E_2((v_{n-3}, f_{n-3}), \ldots, (v_1, f_1)) \end{cases}$$

Now, based on theorem 1, subtask $(T_1, K_1)$ can be performed with only two voltage–frequencies with less energy consumption, i.e.

$$E_1((v_n, f_n), (v_{n-1}, f_{n-1}), (v_{n-2}, f_{n-2})) > \begin{cases} E^*((v_n, f_n), (v_{n-1}, f_{n-1})) \\ or \\ E^*((v_{n-1}, f_{n-1}), (v_{n-2}, f_{n-2})) \\ or \\ E^*((v_n, f_n), (v_{n-2}, f_{n-2})) \end{cases}$$

Thus:

$$E = E_1 + E_2 > E_2 + \begin{cases} E^*((v_n, f_n), (v_{n-1}, f_{n-1})) \\ or \\ E^*((v_{n-1}, f_{n-1}), (v_{n-2}, f_{n-2})) \\ or \\ E^*((v_n, f_n), (v_{n-2}, f_{n-2})) \end{cases}$$

This, in fact, contradicts with the optimality of $E((v_n, f_n),...,(v_1, f_1))$; and therefore, the optimal answer for equation (4) cannot use more than two voltage-frequencies to minimize energy consumption.

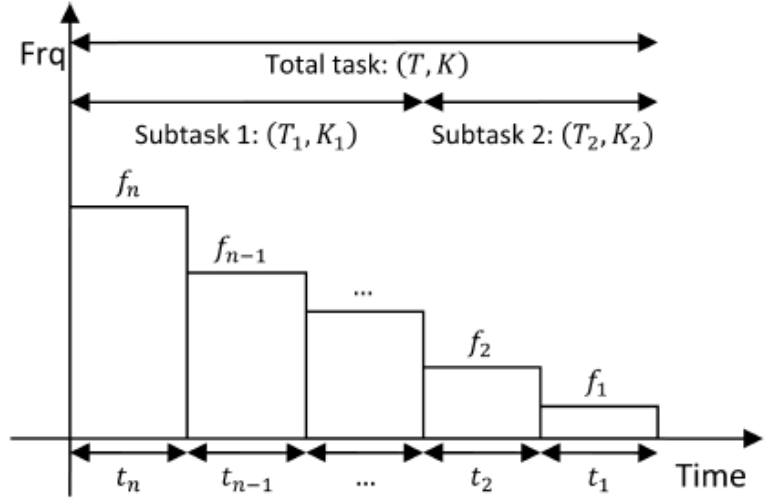

**Figure 2. Optimal answer for equation 4 using multiple frequencies**

Up until now, we managed to prove that equation (4) can only be minimized by using two voltage–frequencies. However, in all these formulas, constraint 2 of this problem was relaxed to use the maximum available time $T$ to find its optimal solution although the optimizer is allowed to use less time than $T$. Therefore, in the following theorem we prove that using less time will always lead to more energy consumption. That is, the original assumption of replacing $t_1 + t_2 + \cdots + t_N \leq T$ with $t_1 + t_2 + \cdots + t_N = T$ was correct.

**Theorem 3:** In equation (4), using less time will always result in consuming more energy.

**Proof:** To prove this theorem, a task is assumed to be executed with two voltage-frequencies $(v_i, f_i)$ and $(v_j, f_j)$ in times $T$ and $T_1(<T)$. Based on corollary 1, associated energy consumption for these two cases would be:

$$T_1: \quad E(T_1) = \frac{T_1 f_j - K}{f_j - f_i} P_d(f_i, v_i) + \frac{K - T_1 f_i}{f_j - f_i} P_d(f_j, v_j) + P_I(T - T_1)$$

$$T: \quad E(T) = \frac{T f_j - K}{f_j - f_i} P_d(f_i, v_i) + \frac{K - T f_i}{f_j - f_i} P_d(f_j, v_j)$$

then,

$$E(T) - E(T_1) = f_j \frac{P_d(f_i, v_i)}{f_j - f_i}(T - T_1) - f_i \frac{P_d(f_j, v_j)}{f_j - f_i}(T - T_1) - P_I(T - T_1)$$

$$= \underbrace{\frac{T - T_1}{f_j - f_i}}_{>0} \left( f_j(P_d(f_i, v_i) - P_I) - f_i(P_d(f_j, v_j) - P_I) \right) \tag{17}$$

As $\frac{f_j}{f_i} > 1$ and $\frac{P_d(f_j, v_j)}{P_d(f_i, v_i)} > 1$, then $\left( f_j(P_d(f_i, v_i) - P_I) - f_i(P_d(f_j, v_j) - P_I) \right) < 0$; and thus, $E(T) - E(T_1) < 0$. Therefore, the original assumption of replacing $t_1 + t_2 + \cdots + t_N \leq T$ with $t_1 + t_2 + \cdots + t_N = T$ is correct.

### 4.3. Computation of Optimal Energy Consumption for $k^{th}$–task

Based on corollary 2, the following post–processing scheduling algorithm is proposed to optimize energy consumption of each task. For $k^{th}$–task, two voltage–frequencies ($(v_i, f_i)$ and $(v_j, f_j)$) that satisfy constraints 1 and 2 from equation (4) (capable of performing $k^{th}$–task in time $T$) are first obtained and then their associated deployment times ($t_i^{(k)}$ and $t_j^{(k)}$) are calculated as follows.

$$\begin{cases} t_i^{(k)} f_i^{(k)} + t_j^{(k)} f_j^{(k)} = K^{(k)} \\ t_i^{(k)} + t_j^{(k)} = T^{(k)} \end{cases} \Rightarrow \begin{cases} t_j^{(k)} = \frac{K^{(k)} - T^{(k)} f_i^{(k)}}{f_j^{(k)} - f_i^{(k)}} \\ t_i^{(k)} = \frac{T^{(k)} f_j^{(k)} - K^{(k)}}{f_j^{(k)} - f_i^{(k)}} \end{cases} \tag{18}$$

Based on constraint 3 from equation (4),

$$\begin{cases} t_j^{(k)} \geq 0 \\ t_i^{(k)} \geq 0 \end{cases} \Rightarrow \begin{cases} K^{(k)} - T^{(k)} f_i^{(k)} \geq 0 \\ T^{(k)} f_j^{(k)} - K^{(k)} \geq 0 \end{cases} \Rightarrow f_i^{(k)} \leq \frac{K^{(k)}}{T^{(k)}} \leq f_j^{(k)}$$

and the optimal energy is calculated as:

$$E^{(k)}(f_i, f_j) = \frac{T^{(k)} f_j - K^{(k)}}{f_j - f_i} P_d(f_i, v_i) + \frac{K^{(k)} - T^{(k)} f_i}{f_j - f_i} P_d(f_j, v_j) \tag{19}$$

Thus, the details of the post–processing algorithm proposed in this paper are as follows:

> **MVFS–DVFS algorithm**
> Post–processing algorithm to optimize energy consumption of scheduled tasks
> - Schedule tasks given by a DAG using a scheduling algorithm
> - **for** k=1:*number of tasks in DAG*
>   - Select the $k^{th}$ task
>   - Calculate $f_{ideal}^{(k)}$
>   - Divide processor frequency set into two groups (*U,L*):
>     $$\underbrace{[f_1,...,f_r]}_{U} > f_{ideal}^{(k)} > \underbrace{[f_{r+1},...,f_N]}_{L}$$
>   - Calculate time and energy from equations (18) and (19) for all $f_j \in U$ and $f_i \in L$
>   - Select $(f_i, f_j)$ associated to the lowest energy for this task
>   **endfor**
> - **return** (individual voltage–frequency pair for execution of each task)

### *4.4. Simplified–Multiple Frequency Selection DVFS (SMFS–DVFS)*

In most of DVFS algorithms, it is assumed that processor energy consumption is a convex function of frequency (or voltage) as:
$$P_d = \lambda f^3$$

The convex function relation between power and voltage was used by Ishihara in [32] where CPU power is just a square function of voltage –not frequency. If the relation between voltage and frequency in equation (1) is assumed to be linear, then the Ishihara work will be similar to the SMFS-DVFS algorithm in this section. Generally, equation (1) is an approximation of the real relation between Voltage-frequency and power in CMOS circuits that may not be followed by a few current or future CPUs. This problem has been solved in MVFS-DVFS algorithm in the previous section of this paper by considering a general form between power and voltage-frequency in CPUs as shown in equation (2). MVFS-DVFS algorithm claims that independent of the way of modelling between power and voltage-frequency, if equation (2) is satisfied, always two frequencies are involved in the optimal energy consumption. In other words, the technique in [32] is a subset of MVFS-DVFS technique described in this paper. This simplification changes the problem statement in equation (4) to:

$$\begin{cases} \text{Minimize} \quad E^{(k)} = \lambda \sum_{i=1}^{N} t_i^{(k)} f_i^3 + P_I \left( T^{(k)} - \sum_{i=1}^{N} t_i^{(k)} \right) \\ \text{s.t.} \\ \quad 1. \quad \sum_{i=1}^{N} t_i^{(k)} f_i = K^{(k)} \\ \quad 2. \quad \sum_{i=1}^{N} t_i^{(k)} \leq T^{(k)} \\ \quad 3. \quad t_i^{(k)} \geq 0; \text{ for } i = 1, 2, \cdots, N \end{cases} \quad (20)$$

Here, to simplify the writing of the equations for the $k^{th}$–task, $t_i^{(k)}$, $T^{(k)}$ and $K^{(k)}$ are also replaced with $t_i$, $T$ and $K$, respectively. As this simplified problem is a case−study of the main problem, all the proved theorems and corollaries are still valid; therefore, (1) two frequencies $f_i$ and $f_j (> f_i)$ result in the optimal answer, and (2) these two frequencies are near $f_{ideal} = \dfrac{K}{T}$ or $f_i < f_{ideal} < f_j$. The following two theorems, exclusively proved for this case study, show that $f_i$ and $f_j$ must also be adjacent. In this case (cubic functions) the optimal result can be calculated as:

$$\begin{aligned} E^*(f_i, f_j) &= \lambda K \frac{f_i^3 - f_j^3}{f_i - f_j} - \lambda T f_i f_j \frac{f_i^2 - f_j^2}{f_i - f_j} \\ &= \lambda \left[ K(f_i^2 + f_i f_j + f_j^2) - T f_i f_j (f_i + f_j) \right] \end{aligned} \quad (21)$$

**Theorem 4:** if $f_i$ and $f_j$ are capable of performing a task and there exists $f_r$ such that $f_i < f_r < f_j$, then adding $f_r$ to the frequency pool will always reduce the total energy consumption.

**Proof:** to prove this theorem we need to prove that:

$$E(f_i, f_j) > \begin{cases} E^*(f_r, f_j) & \text{if } K - T f_r > 0 \\ E^*(f_i, f_r) & \text{if } K - T f_r < 0 \end{cases}$$

**Case 1: IF** $K - T f_r \geq 0$:

$$\begin{aligned} E(f_i, f_j) - E^*(f_r, f_j) &= \\ &= \left[ K(f_i^2 + f_i f_j + f_j^2) - T f_i f_j (f_i + f_j) \right] - \left[ K(f_r^2 + f_r f_j + f_j^2) - T f_r f_j (f_r + f_j) \right] \\ &= \ldots \\ &= (f_r - f_i)(f_i + f_r + f_j)(T f_j - K) \\ &= (f_r - f_i)(f_i + f_r + f_j)(f_j - f_r) f_r \\ &> 0 \end{aligned}$$

**Case 2: IF** $K - Tf_r < 0$:

$$E(f_i, f_j) - E^*(f_1, f_r) =$$
$$= [K(f_i^2 + f_i f_j + f_j^2) - Tf_i f_j (f_i + f_j)] - [K(f_i^2 + f_i f_r + f_r^2) - Tf_i f_r (f_i + f_r)]$$
$$= \ldots$$
$$= (f_j - f_r)(f_i + f_r + f_j)(K - Tf_i)$$
$$= (f_j - f_r)(f_i + f_r + f_j)(f_r - f_i) t_r$$
$$> 0$$

Using the above mentioned theorems, the optimal answer for equation (20) can now be calculated. Theorem 2 proves that regardless of the number of available frequencies for a processor, the optimal answer would use at most two frequencies, while theorem 5 proves that these two frequencies should be also adjacent.

**Theorem 5:** The two frequencies that minimize equation (20) are adjacent.

**Proof (by contradiction):** Based on theorem 2, the optimal answer for equation (20) can be only obtained by using two frequencies. Here, we prove that these two frequencies should also be adjacent. To prove that (by contradiction), we show that these two frequencies cannot be non–adjacent. If there are, as shown in Figure 3, theorem 4 suggests that adding any available frequency between these two frequencies will reduce the total energy consumption and yields another answer with less energy consumption. This, in fact, contradicts the optimality of the original non–adjacent frequency selection. Therefore, two frequencies that minimize equation (20) should be adjacent.

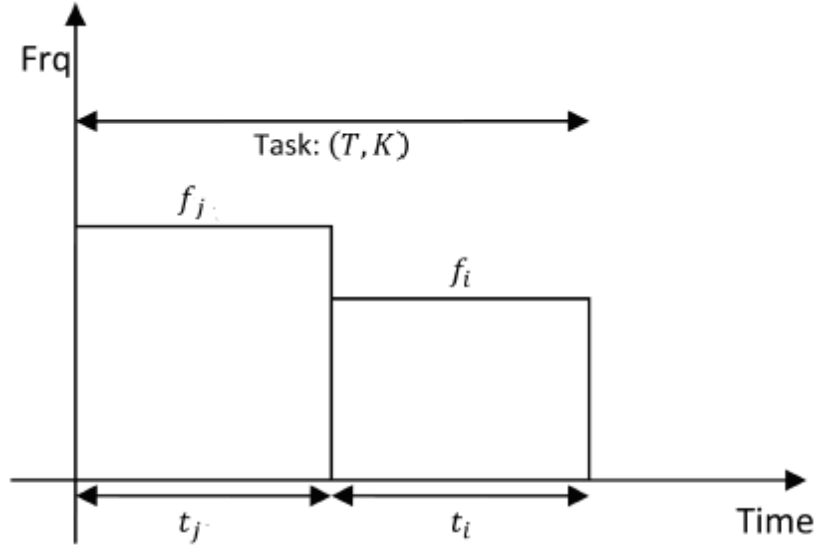

**Figure 3. Optimal answer for equation 4 with two voltage–frequencies**

So far, we proved that equation (20) can only be minimized by using two adjacent frequencies. As the only adjacent frequencies before and after $f_{ideal} = \frac{K}{T}$ are $f_{RD}$ and $f_{RD-1}(< f_{RD})$, respectively, therefore the optimal energy is achieved by frequencies $(f_{RD}, f_{RD-1})$, where $f_{RD}$ is a frequency obtained from the RDVFS algorithm in [11] and is defined as the largest and closest frequency to $f_{ideal} = \frac{K}{T}$. For $k^{th}$–task, the associated times for $f_{RD}$ and $f_{RD-1}$ are calculated as follows:

$$\begin{cases} t_{f_{RD-1}} = \dfrac{Tf_{RD} - K}{f_{RD} - f_{RD-1}} \\ t_{f_{RD}} = \dfrac{K - Tf_{RD-1}}{f_{RD} - f_{RD-1}} \end{cases} \quad (22)$$

Therefore the algorithm for SMFS−DVFS will be:

---

**SMFS-DVFS algorithm**
<u>Post-processing algorithm to optimally energy consumption of scheduled tasks</u>
- Schedule tasks given by a DAG using an scheduling algorithm
- **for** k=1:*number of tasks in DAG*
  - Select the k$^{th}$ task
  - Calculate $f_{ideal}^{(k)}$
  - Select the immediate frequencies in the processor frequency set before and after $f_{ideal}^{(k)}$. These frequencies are $f_{RD}^{(k)}$ and $f_{RD-1}^{(k)}$.
  - Calculate associated times from equation (22) and energy of task from equation (21).
  - Select $(f_{RD}^{(k)}, f_{RD-1}^{(k)})$ associated to the lowest energy for this task
  **Endfor**
- **return** (individual frequencies pair for execution of each task)

---

## 5. EXPERIMENTAL RESULTS AND DISCUSSION

This section presents simulation results of our proposed algorithm (MVFS−DVFS) as well as other algorithms (RDVFS, MMF−DVFS and optimum continuous frequency) for a more comprehensive comparison. Here, the following three schedulers are used to produce original task schedules: (1) list scheduling, (2) list scheduling with Longest Processing Time first (LPT) and (3) list scheduling with Shortest Processing Time first (SPT) are employed with different numbers of processors. The simulator itself was developed as a part of this study.

### *5.1 An Example*

The following example shows how each of the algorithms uses a task's slack time to reduce its associated energy consumption. To simplify, it is assumed that the power consumption is a cubic function of frequency as $p_d(v,f) = 1.367 \times 10^{-24} f^3$. Figure 1a shows the original scheduling of $k^{th}$–task ($A^{(k)}$) executed on a processor. Assuming $P_{Idle} = 0$, the values of the parameters for this task are as follows:

| | |
|---|---|
| $f_{RD}^{(k)}$ | 60MHz |
| $f_{RD-1}^{(k)}$ | 50MHz |
| $t_{OS}^{(k)}$ | 70 msec |
| $T^{(k)}$ | 130 msec |
| $K^{(k)}$ | 7 million cycles |

Based on these parameters:

- By referring to equations (14) and (15), the optimum continuous frequency is calculated as $f_{ideal}^{(k)} = \frac{K^{(k)}}{T^{(k)}} = 53.84 MHz$, which is not a valid frequency in the processor frequency list. The energy corresponding to this frequency is $E_{Opt-Cont.}^{(k)} = 27.73\ mW$ (Figure 1c). As we mentioned earlier, this energy is the optimum energy for this task.
- In the RDVFS algorithm, this processor executes the task with the most available frequency close to the $f_{ideal}^{(k)}$. Refer to the previous table, this frequency ($f_{RD}^{(k)}$) is 60*MHz*. Referring to equation (3), the energy calculated by this method is $E_{RD}^{(k)} = 34.25\ mW$ (Figure 1b).
- SMFS–DVFS (the simplified version of our proposed method) attempts to find the optimal energy by a linear combination of all processor frequencies. We proved that for each task always two neighbour frequencies produce the optimal energy. These two frequencies are $f_{RD}^{(k)}$ and $f_{RD-1}^{(k)} (< f_{RD}^{(k)})$ which are obtained from RDVFS algorithm (Figure 1d). The value of energy consumption of $k^{th}$ task in this example is calculated as $E_{MFS-DVFS}^{(k)} = 28.43\ mW$. It can be noted that SMFS–DVFS gives the closest energy to the optimum energy ($E_{Opt-Cont.}^{(k)}$) compared with RDVFS algorithm.

## *5.2. Experimental Settings*

### *5.2.1. Processor models*

Voltage/frequency settings are defined based on two groups of processors: the first group includes two synthetic processors, while the second group includes two real processors (Transmeta Crusoe [35] and Intel Xscale [36]). Table 2 shows the voltage/frequency and the related power consumption of these processors.

*5.2.2. Task information*

The performance of MVFS–DVFS was evaluated with two sets of task graphs: randomly generated and real–world applications. For each application, a large number of variations in the number of tasks and the number of processors were applied to simulations.

The random task graphs set consists of 3000 graphs with five graph sizes, three different schedulers and five sets of processors. These task graphs have different number of tasks, task distributions, communication costs and task dependencies. The execution cycle of these randomly generated tasks varied from 5–10 million cycles from a uniform distribution, respectively. The two applications used in these experiments were LU decomposition and Gauss–Jordan with directed acyclic graph (DAG) and execution cycles extracted from[19]. Also, 600 real–world application task graphs based on Gauss–Jordan and LU decomposition algorithms were used in the experiments. For each application graph, the same number of task graphs (ranging from 100 to 500 tasks) with three schedulers and on five sets of processors were investigated.

**Table 1. Experimental parameters**

| Parameter | Value |
|---|---|
| The number of tasks | [100, 200, 300, 400, 500] |
| The number of processors in clusters | [2, 4, 8, 16, 32] |
| Processor type | 2 synthetic processor, Transmeta Crusoe, Intel Xscale |

**Table 2. Voltage/frequency settings of four processors with their associated power consumption**

| | Synthetic Processor 1 | | | Synthetic Processor 2 | | |
|---|---|---|---|---|---|---|
| *Level* | *Frequency (MHz)* | *Voltage (V)* | *Power (W)* | *Frequency (MHz)* | *Voltage (V)* | *Power (W)* |
| 0 | 1000 | 1.2 | 7.2 | 1000 | 1.25 | 5.0 |
| 1 | 900 | 1.15 | 5.95 | 900 | 1.05 | 3.29 |
| 2 | 800 | 1.1 | 4.84 | 500 | 1.00 | 2.05 |
| 3 | 700 | 1.05 | 3.85 | 400 | 0.95 | 1.64 |
| 4 | 600 | 1.00 | 3 | 300 | 0.90 | 0.97 |
| 5 | 500 | 0.9 | 2.03 | --- | --- | --- |

| | Transmeta Crusoe [35] | | | Intel Xscale [36] | | |
|---|---|---|---|---|---|---|
| *Level* | *Frequency (MHz)* | *Voltage (V)* | *Power (W)* | *Frequency (MHz)* | *Voltage (V)* | *Power (W)* |
| 0 | 667 | 1.6 | 5.3 | 1000 | 1.8 | 1.6 |
| 1 | 600 | 1.5 | 4.2 | 800 | 1.6 | 0.9 |
| 2 | 533 | 1.35 | 3.0 | 600 | 1.3 | 0.4 |
| 3 | 400 | 1.225 | 1.9 | 400 | 1 | 0.17 |
| 4 | 300 | 1.2 | 1.3 | 150 | 0.75 | .08 |

### 5.3. Results and Discussions

Table 3 shows the simulation results of normalized energy consumption for all DAG sets (Figures 4 and 5). This table clearly indicates the superior performance of MVFS–DVFS compared with others in all cases. The performance of MVFS–DVFS and other related algorithms has a strong dependency on tasks' slack times in the original scheduling. This dependency explains why the algorithms are not successful in reducing the energy consumption of Gauss–Jordan task graphs. To clarify, a three level Gauss–Jordan task scheduling on three homogenous processors is shown in Figure 6, which clearly shows that the relations among tasks and their computation and communication costs leave no slack time for tasks to be used by the algorithms.

**Table 3. The energy saving percentage of MVFS–DVFS and other related algorithms on 3600 random and real graphs**

| Experiment | RDVFS | MMF-DVFS | MVFS-DVFS | Optimum Continuous frequency |
|---|---|---|---|---|
| Random tasks | 13.00% | 13.50% | 14.40% | 14.84% |
| Gauss–Jordan | 0.1% | 0.11% | 0.11% | 0.14% |
| LU–decomposition | 24.8% | 25.5% | 27.0% | 27.81% |

Besides the effectiveness of MVFS-DVFS compared with other slack reclamation algorithms, some other issues should be addressed. The first issue is the relationship between energy consumption and the number of processors in our experiments. Increasing the number of processors expedites the processing time and therefore reduces the makespan; as a side–effect however, it also increases system's slack times. Figure 7, addressing this effect, shows the percentage of overall energy saving for a system with different number of processors for random and LU decomposition task graphs. This figure indicates that for both random and LU decomposition task graphs increasing the number of processors results in saving more energy by these algorithms. This figure also shows the influence of the type of scheduling on random task graphs which results in increasing the amount of slack time between tasks for 8 and 16 processors compare with 4 and 32 processors.

The second issue which is the major limitation on most DVFS–based algorithms working with one frequency (such as the RDVFS algorithm) is that the slack time can not been covered by using only one frequency. Those algorithms work better when processors can be used at any arbitrary frequency. Nevertheless, due to technological issues, the number of valid frequencies is limited; therefore, these algorithms must select the most suitable frequency among a set of frequencies, defined by DVFS, instead. Generally, the relation among $t_{RD}^{(k)}$, $f_{RD}^{(k)}$, $f_{\max}$ and $t_{OS}^{(k)}$ for task $A^{(k)}$ is: $t_{RD}^{(k)} = \dfrac{f_{RD}^{(k)}}{f_{\max}} t_{OS}^{(k)}$ (suppose the processor works in $f_{\max}$ in the original scheduling). Thus, although $t_{RD}^{(k)}$ is a continues–variable, it cannot accept any arbitrary value; therefore, the slack time of tasks cannot be minimized. On the other hand, in SMFS–DVFS algorithm, the relation between those variables is: $f_{RD}^{(k)} t_{RD}^{(k)} = f_{RD}^{(k)} t_{f_{RD}}^{(k)} + f_{RD-1}^{(k)} t_{f_{RD-1}}^{(k)}$, an equation with two variables ($t_{f_{RD}}^{(k)}, t_{f_{RD-1}}^{(k)}$) that might have many eligible answers. Thus, appropriate values of these variables, in relation to task conditions, can minimize the slack time and/or reduce energy consumption.

The third issue is the overhead of running MVFS–DVFS. This overhead comes from the transition time of switching from one frequency to another frequency. An almost true assumption is that the overhead of transition times are relatively much less than the execution times of tasks; therefore the transition times overhead can be neglected in calculations. In our experiments, tasks with duration at least 100 times longer than their transition times are considered for the MVFS–DVFS algorithm.

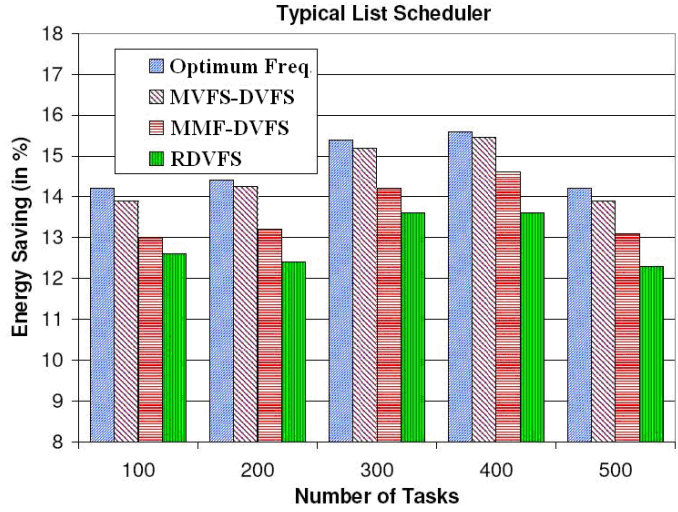

(a)

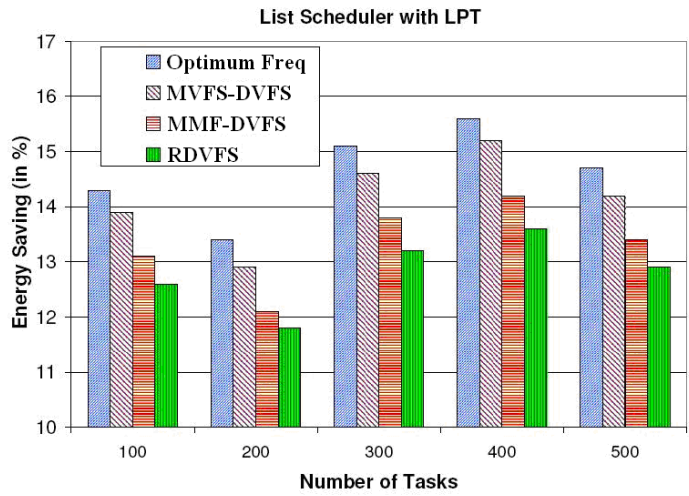

(b)

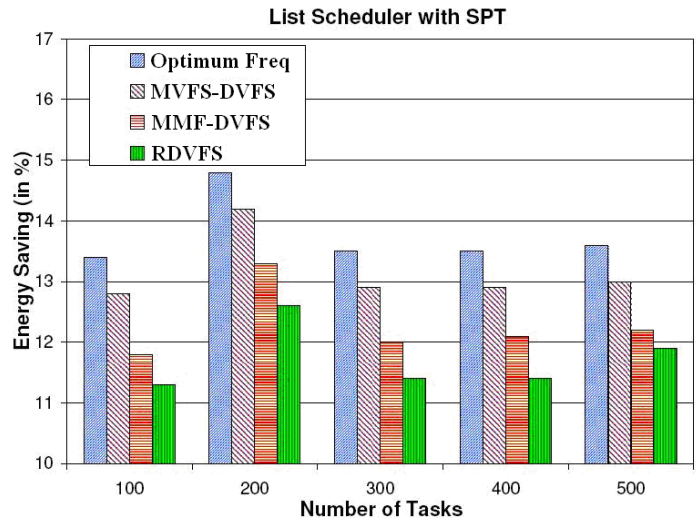

(c)

**Figure 4. The normalized energy saving on the number of tasks for MVFS–DVFS algorithm compared with other algorithms for three list schedulers: (a) The typical list scheduler (b) The list scheduler with Longest Processing Time first (LPT) and (c) The list scheduler with Shortest Processing Time first (SPT). The tasks in this figure are generated and averaged based on 3000 random experiments for 2, 4, 8, 16 and 32 processors.**

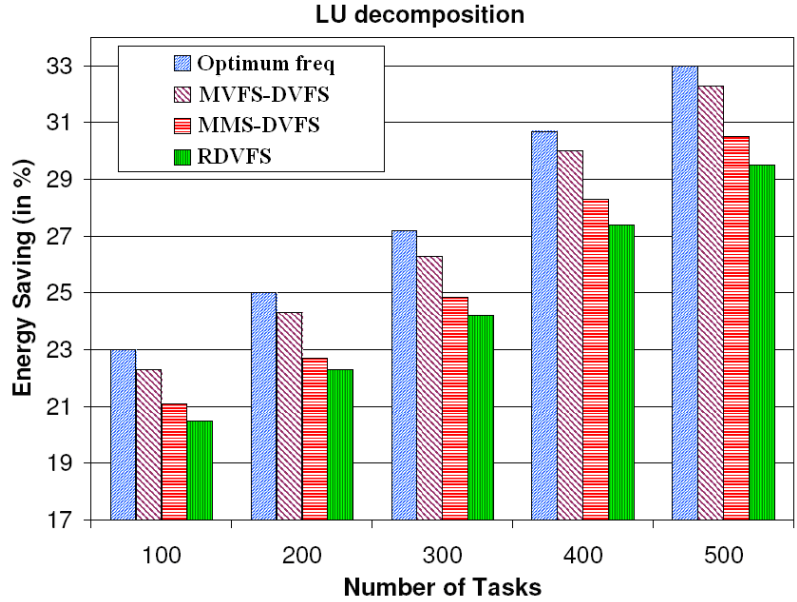

(a)

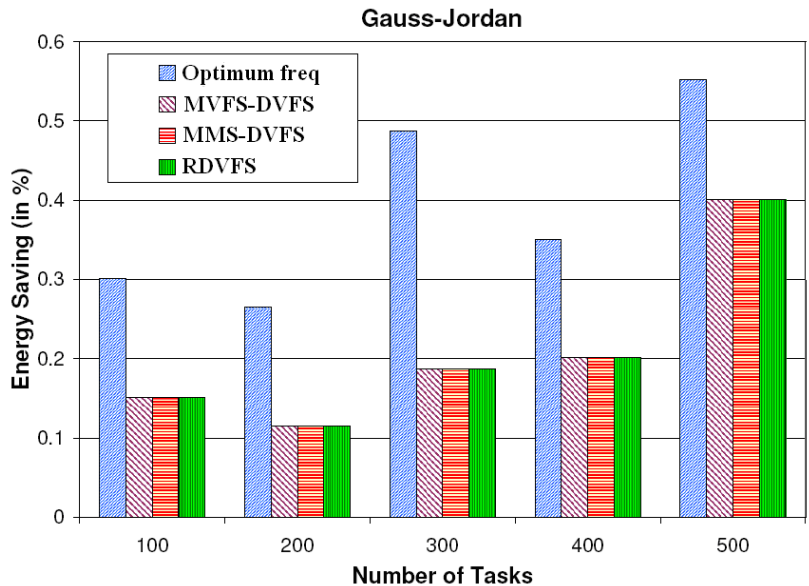

(b)

**Figure 5. The normalized energy saving of MVFS–DVFS and other algorithms on the number of tasks for two real-world applications: (a) Gauss-Jordan, (b) LU decomposition. The tasks are generated and averaged based on 600 experiments; all were scheduled by list scheduler with Longest Processing Time first (LPT) for 2, 4, 8, 16 and 32 processors.**

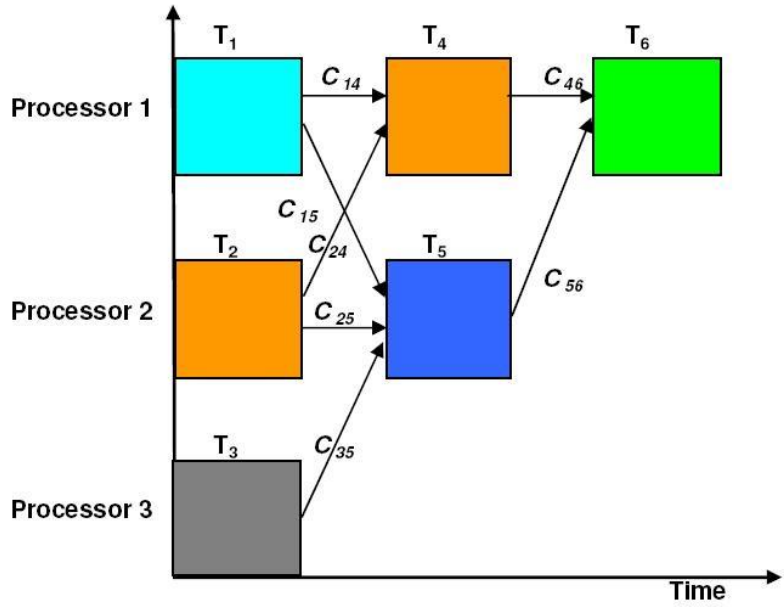

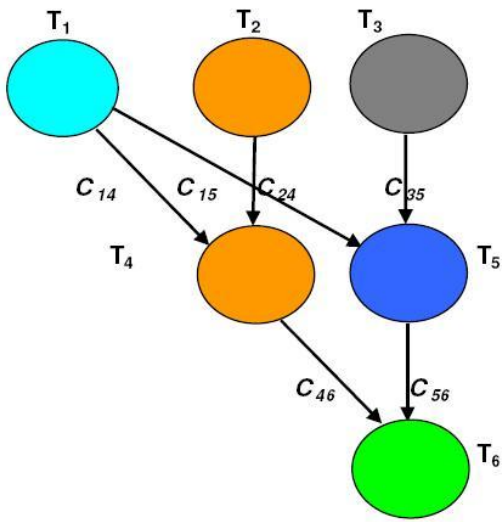

**Figure 6. Gauss–Jordan task graph: (a) a sample scheduling of a three level Gauss–Jordan task graph on three processors, (b) a Gauss-Jordan DAG for three levels. The communication costs ($C_{ij}$) are equal to 10 time units for all *i* and *j*.**

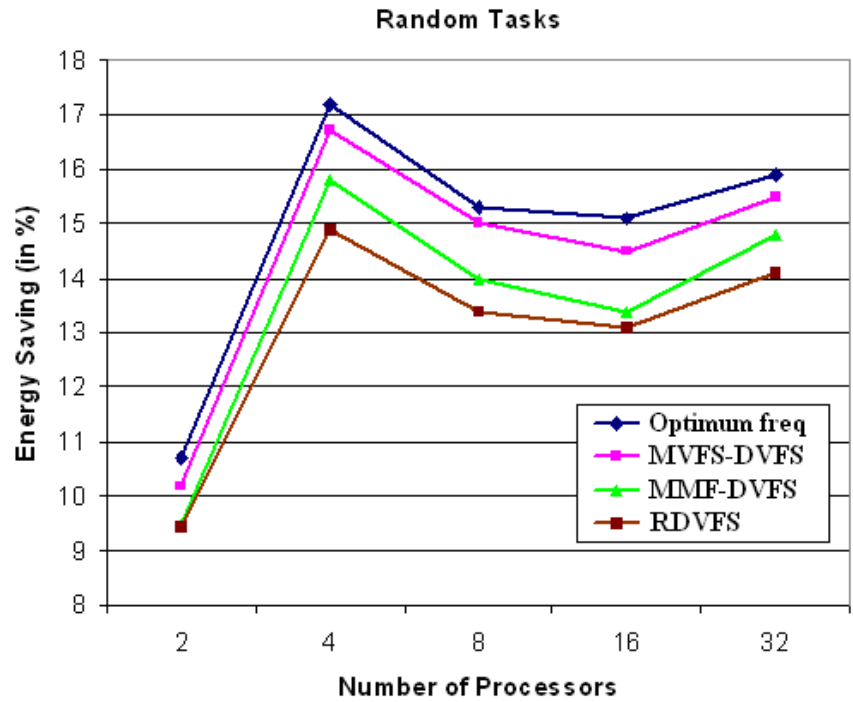

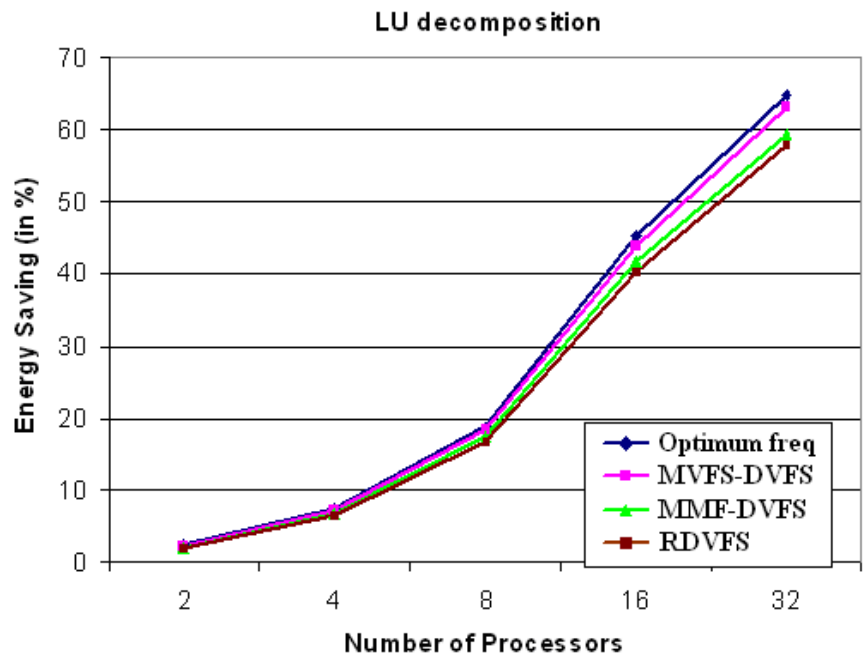

**Figure 7. The comparison between the percentage of energy saving in MVFS–DVFS and other algorithms on the number of processors: (a) 3000 randomly generated task graphs, (b) 300 LU decomposition task graphs.**

# 6. Conclusion

Since most traditional static task scheduling algorithms for HPC systems do not consider power management, we addressed the energy issue with task scheduling in clusters and presented the MVFS–DVFS algorithm which is based on the DVFS technique. In this work, we specifically studied the use of a linear combination of more than one voltage-frequency to reduce energy consumption on processors. We proved that the optimal energy in a discrete set of voltage-frequencies for each task is achieved by a combination of two voltage-frequencies. These two voltage-frequencies are adjacent when the power consumption of the processor is a convex function of frequency. Simulation results of 3000 randomly generated task graphs and 600 real application task graphs showed the effectiveness of the MVFS–DVFS algorithm compared with other related algorithms. The MVFS–DVFS consumes the least amount of energy of all cases.

# 7. Acknowledgment

The work reported in this paper is in part supported by National ICT Australia (NICTA). Professor A.Y. Zomaya's work is supported by an Australian Research Council Grant DP1097110.